\documentclass[12pt]{iopart}

\usepackage{graphicx,type1cm,eso-pic,color}
\usepackage{setspace}

\newcommand{\mg}{$^{25}$Mg$^+$}
\newcommand{\mindex}[1]{\textrm{\tiny #1}}
\newcommand{\ket}[1]{|#1\rangle}

\newcommand{\upstate}{$\ket{\!\uparrow}$}
\newcommand{\downstate}{$\ket{\!\downarrow}$}

\newcommand{\rfig}[1]{Figure \ref{#1}}
\newcommand{\rsec}[1]{Section \ref{#1}}
\newcommand{\rapp}[1]{\ref{#1}}
\newcommand{\req}[1]{Eq.~\ref{#1}}
\newcommand{\ind}[1]{\textrm{\tiny #1}}

\usepackage{mathptmx} 

\begin{document}
\title[Quantum State Detection Schemes]{A Novel, Robust Quantum Detection Scheme}

\author{B Hemmerling$^{1,4}$ and F Gebert$^1$ and Y Wan$^{1,3}$ and P O Schmidt$^{1,2}$}
\address{$^1$ QUEST Institute for Experimental Quantum Metrology, Physikalisch-Technische Bundesanstalt, 38116 Braunschweig, Germany}
\address{$^2$ Institut f\"ur Quantenoptik, Leibniz Universit\"at Hannover, 30167 Hannover, Germany}
\address{$^3$ Braunschweig International Graduate School of Metrology, Technische Universit\"at Braunschweig, 38106 Braunschweig, Germany}
\address{$^4$ {\it Present address:} Department of Physics, Harvard University, Cambridge, Massachusetts 02138, USA}
\ead{boerge.hemmerling@quantummetrology.de}

\begin{abstract}
Protocols used in quantum information and precision spectroscopy rely on efficient internal quantum state discrimination. With a single ion in a linear Paul trap, we implement a novel detection method which utilizes correlations between two detection events with an intermediate spin-flip. The technique is experimentally characterized to be more robust against fluctuations in detection laser power compared to conventionally implemented methods. Furthermore, systematic detection errors which limit the Rabi oscillation contrast in conventional methods are overcome.
\end{abstract}

\maketitle

\section{Introduction}

A standard protocol for quantum mechanics experiments is comprised of a three-step sequence: Firstly, the quantum system is initialized to a known quantum state, for instance by optical pumping in the case of atoms or ions. Secondly, external control fields, e.g.~sequences of laser light pulses or radio-frequency radiation, are applied to implement the Hamiltonian of interest. Finally, the resulting state is read out by a projective measurement. In the case of single or few-atom quantum systems, repeating the experiment many times is necessary in order to accumulate statistics, such that the average measurement result represents the quantum state after the evolution according to the Hamiltonian, projected onto the measurement basis. Thus, the fidelity of state detection plays an essential role in the understanding of the underlying dynamics and, in addition to other constraints given by the experimental setup, imposes a limitation to the quality of the experimental findings.
 
Trapped ions are an ideal system to investigate quantum mechanical effects owing to the almost perfect control of the ion's internal and motional state preparation, manipulation and read-out \cite{Zoller:95}. This is demonstrated by both relevant theoretical and experimental progress, providing new insights in many fields of physics, to name a few: quantum computing \cite{Blatt:08}, optical clocks \cite{Chou:10,Rosenband:08}, precision spectroscopy \cite{Schmidt:05,Roos:06,Wolf:09} and quantum simulations \cite{Wineland:02,Friedenauer:08,Gerritsma:10,Kim:10}. All underlying experimental schemes depend on the high fidelity of both the applied algorithm and the state detection of the ion.

The basic principle of state discrimination in ions relies on {\it electron-shelving} \cite{Dehmelt:75b}. Two different energy states (qubit) are distinguished by their state-dependent fluorescence via coupling to a third level. In its simplest form, the number of collected photons during a single detection cycle determines whether the ion is assigned to a so-called bright (dark) state depending on this number being higher (lower) than a chosen threshold. This {\it threshold detection technique} has been successfully implemented in many experiments yielding an almost unity read-out fidelity for qubits stored in two optically separated meta-stable states ($^{40}$Ca$^+$ \cite{Steane:08}, $^{111}$Cd$^+$ \cite{Monroe:05}, $^{88}$Sr$^+$ \cite{Keselman:11}). Detection fidelities can be further improved by taking into account photon arrival times, as opposed to only observing the integrated signal. This type of {\it Bayesian inference} or {\it maximum likelihood detection} has been successfully demonstrated in an optical qubit with $^{40}$Ca$^+$ yielding fidelities of 99.991(1)\% \cite{Steane:08, Lucas:09}. The detection fidelity is ultimately limited by the finite overlap between the bright and dark state photon distributions, resulting in an increased and asymmetric (bias) error in the assignment of qubit states. Long detection times reduce this error, but are ultimately limited in real physical systems by the finite lifetime of the qubit states due to spontaneous or induced transitions to other states. This effect is particularly prominent for hyperfine qubits, where the detection fidelity suffers from depumping of one of the qubit states ($^{9}$Be$^+$ \cite{Wineland:97, Langer:00}, $^{171}$Yb$^+$ \cite{Monroe:07}, \mg\ \cite{Hemmerling:11}), effectively limiting the detection time and thus the number of observed photons. This can be overcome by implementing a quantum non-demolition measurement in which the qubit state to be detected is repeatedly transferred to an auxiliary ion, where detection is performed. A Bayesian state inference of the time series of photon detection events of such an experiment, implemented with an optical qubit in Al$^+$, and read out by a hyperfine qubit in Be$^+$, yielded a detection fidelity of 99.94\% \cite{Wineland:07b}.
However, achieving such a high fidelity requires either an auxiliary ion to implement a quantum non-demolition measurement \cite{Wineland:07b}, or an optical qubit with a long excited state lifetime. Here, we demonstrate a novel state detection technique with improved state discrimination by combining two detection events of either threshold or Bayesian detection types with an intermediate well-controlled state inversion spin-flip. Observation of the correlated detection outcome acts as a post-selective statistical filter, significantly reducing the state assignment errors due to overlapping photon distributions. Owing to the symmetry of the $\pi$-detection scheme, the detection errors for the two qubit states are at the same time equalized, effectively reducing the bias error. Another implication of this effect is the robustness of the scheme against fluctuations of detection parameters, making it particularly well-suited for systems in which only very few photons are detected. Consequently, the $\pi$-detection scheme will always improve the detection fidelity of threshold and Bayesian detection schemes if the state inversion can be performed with sufficient fidelity. The scheme is also applicable to multi-particle systems, as long as individual atoms can be resolved.
Furthermore, we imagine similar schemes to be applied to neutral atoms with non-destructive detection methods where the particle is not lost after a single detection cycle, such as measuring the state-dependent reflection and transmission signals of molecules or atoms in cavities \cite{Zoller:06, Reichel:10}.

\section{Experimental Apparatus}

The experiments discussed here have been carried out in a setup described in detail in \cite{Hemmerling:11}. Briefly, we trap a single \mg\ ion in a linear Paul trap at trapping frequencies $\omega_{\mindex{ax}} \sim 2\pi \times 2.2$\,MHz in the axial and $\omega_{\mindex{r}} \sim 2\pi \times 4.5$\,MHz in both radial directions. The qubit is encoded into two long-lived hyperfine ground states of the S$_{1/2}$ manifold, namely $\ket{F,m_F} = \ket{3,3} = \ket{\!\downarrow}$ and $\ket{2,2} = \ket{\!\uparrow}$, representing the bright and the dark state, respectively. For all measurements, a magnetic field of typically $B \sim 0.6$\,mT lifts the degeneracy between the magnetic sub-levels. A frequency-quadrupled solid-state fiber laser system provides light at 280\,nm for laser cooling and detection using the cycling transition between the \downstate\ and the $\ket{4,4} = \ket{e}$ state of the P$_{3/2}$ manifold with a linewidth of $\Gamma \sim 2\pi \times 40$\,MHz. The detection and Doppler cooling lasers are derived from the same source, their only difference being that the detection laser operates on resonance with the cycling transition, whereas the Doppler cooling laser is detuned by half the natural linewidth for optimal cooling performance. 

\begin{figure}[!h]\begin{center}
	\includegraphics[width=0.75\linewidth]{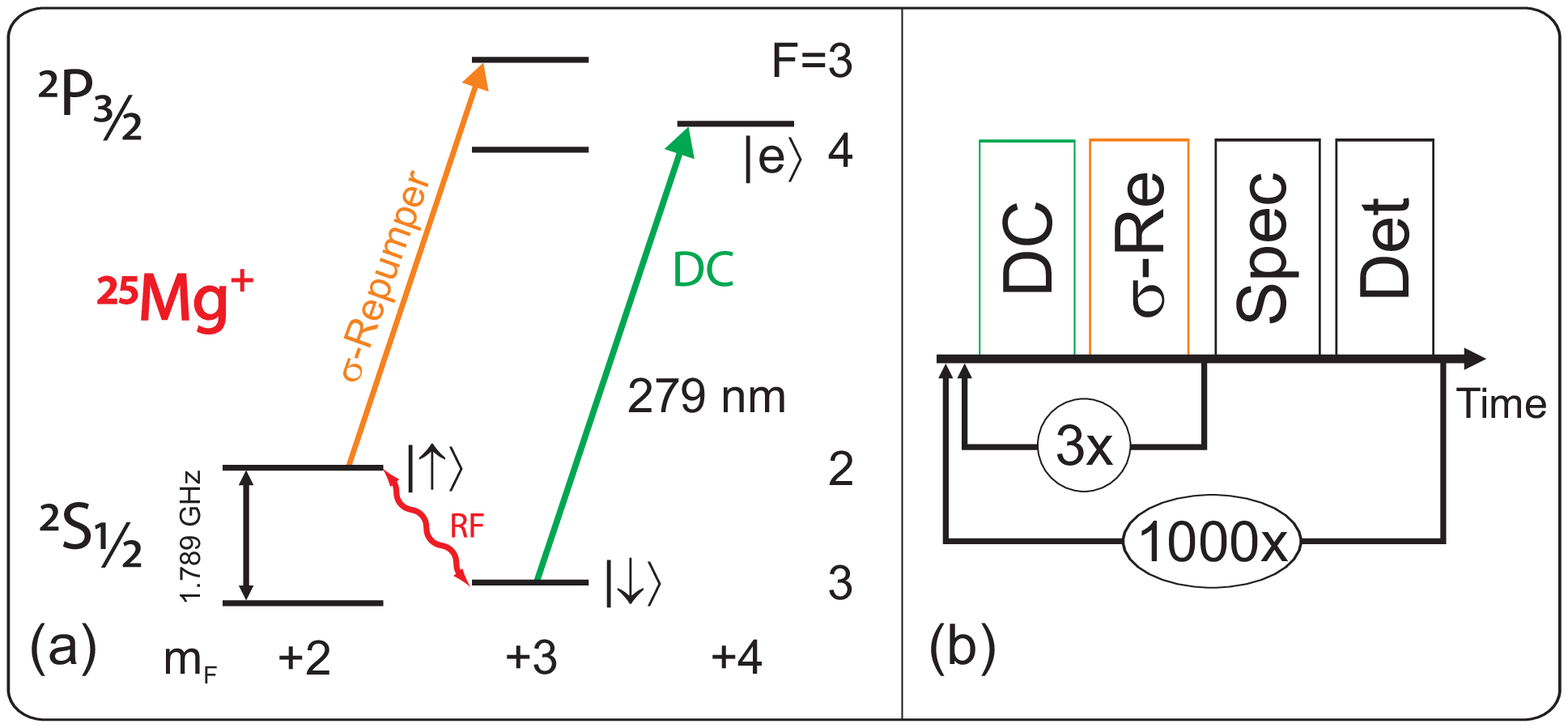}
	\caption{
\label{fig:level_scheme}
(a) Magnesium level scheme. Only relevant levels and transitions are shown. Doppler cooling ({\it DC}) is applied on the cycling transition between the S$_{1/2}$ and P$_{3/2}$ manifolds. An additional {\it $\sigma$-repumper} improves state initialization. The qubit levels \downstate\ and \upstate\ are coherently coupled by applying radio-frequency pulses ({\it RF}) at 1.789\,GHz.
(b) Experimental protocol. Three Doppler cooling and repumping pulses initialize the ion to the \downstate\ state. After that, a radio-frequency spectroscopy pulse is applied. The sequence is concluded with a resonant detection pulse (same as the {\it DC} beam). To obtain sufficient statistics, the sequence is typically repeated for 300-1000 times.
}
\end{center}\end{figure}

The resulting resonance fluorescence scattered from the ion is collected by both an objective and a parabolic mirror and subsequently focused onto a photo-multiplier tube for quantitative readout on a $\mu$s time scale. The application of radio-frequency at 1.789\,GHz, which corresponds to the hyperfine splitting in \mg, provides coherent coupling between the qubit states. The radio-frequency is supplied by a quarter-wave antenna at a distance of approximately 12\,cm from the center of the ion trap. An overview of all relevant levels and transitions is shown in \rfig{fig:level_scheme}\,(a).

\section{Experimental Protocol and Quantum State Detection}

After the ion is loaded into the trap, the experimental sequence comprises three steps (see \rfig{fig:level_scheme}(b)) which are repeatedly applied to the ion:

{\bf 1. Preparation and Cooling.} The ion is initialized by optical pumping to \downstate\ by means of a Doppler cooling laser which is applied for 600\,$\mu$s. An additional repumping laser guarantees that no population remains in the upper manifold (S$_{1/2} \,F=2$). After Doppler cooling, the temperature of the ion is close to the Doppler limit of $\sim\!1$\,mK \cite{Hemmerling:11}.

{\bf 2. Radio-Frequency Spectroscopy.} A radio-frequency spectroscopy pulse is applied for a time $\tau_\mindex{spec}$ to drive magnetic field induced Rabi oscillations between \downstate\ and \upstate. Due to the small Lamb-Dicke parameter of the radio-frequency transition $\eta \sim\!10^{-7}$, the transition is independent of the motional state of the ion, which makes Doppler cooling sufficient for these experiments.

{\bf 3. State Detection.} The quantum state of the ion is read-out using different protocols, involving at least one detection laser pulse on resonance with the $\ket{\downarrow} \leftrightarrow \ket{e}$ transition. At the end of each experimental cycle the ion is projected into either the \downstate\ or the \upstate\ state. The ion's state is inferred by averaging over the results of many identical experimental cycles.

If the ion is in the \downstate\ state, application of the detection laser at saturation intensity will let it reside in the cycling transition and scatter photons with a rate of approximately $\Gamma/4 \sim\!6.425 \times 10^7$\,s$^{-1}$, where $\Gamma$ corresponds to the natural linewidth of the magnesium transition \cite{Nist}. On the other hand, if the ion is in the \upstate\ state , the detection laser is detuned by the ground state splitting of $\Delta_\mindex{HF} \approx 2\pi \times 1.789$\,GHz and photon scattering is suppressed by $\Gamma^2/2\Delta^2$.

The collected number of photons are random variables. For the bright state, the random variable follows a Poissonian distribution \cite{LoudonBook}. The probability to measure $k$ photons is given by
\begin{eqnarray}
p( \xi = k | \bar{\xi}) = \frac{e^{-\bar{\xi}} \cdot \bar{\xi}^k}{k!}\quad,
\end{eqnarray}
where $\bar{\xi} = R \cdot \tau$ denotes the average photon number, $R$ is the scattering rate and $\tau$ is the detection time.

\subsection{State Detection Errors}

There are several sources of systematic errors which yield false state information. They impose limits on the detection time and the read-out fidelity. In the following, these errors are briefly described. Afterwards, different detection methods which help overcome these limitations are discussed. A detailed theoretical discussion on detection errors of the threshold and Bayesian methods is found in \cite{Schoelkopf:07}.

\subsubsection{Depumping and Decay of the Upper Qubit State.}

\begin{figure}[!h]\begin{center}
	\includegraphics[width=1.0\linewidth]{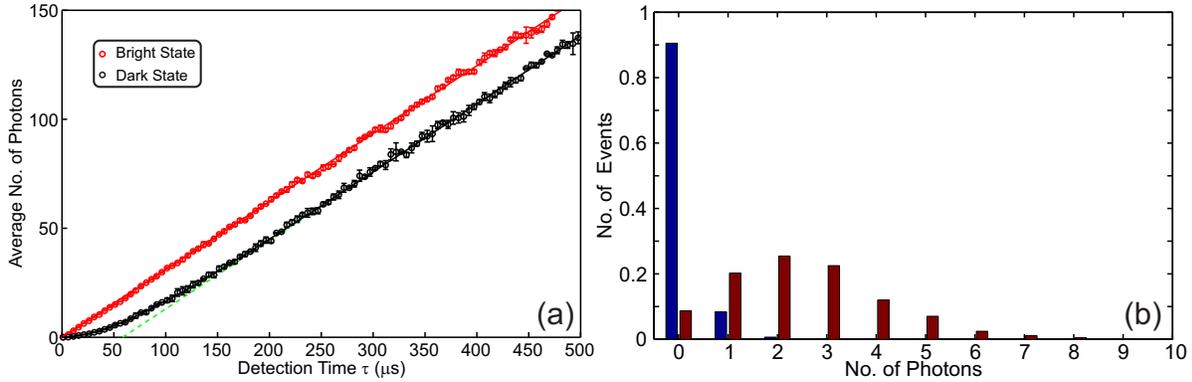}
	\caption{
\label{fig:depumping}
(a) Depumping of dark state. The measured average number of collected photons as a function of the detection time is shown. The photon number increases linearly, if the ion is in the bright state ({\it red circles}). The {\it solid red line} corresponds to a fit to \req{eq:bright_fluorescence}. For an ion in the dark state ({\it black circles}), the count rate reflects only the background photon counts for short times. The detection laser eventually depumps the ion to the bright state, thus the count rate asymptotically approaches the rate of the bright state for long detection times. The {\it solid black line} corresponds to a fit to \req{eq:dark_fluorescence} and yields a decay time of $T \approx 61 \pm 4\,\mu$s. The {\it dashed green line} reflects the asymptotic behaviour of the dark fluorescence for large detection times.
(b) Measured normalized example photon histograms of the bright and the dark state for a detection time of 10\,$\mu$s for a single \mg\ ion. The measured scattering rates for the bright (dark) state are $R_\ind{B} \sim 249 \cdot 10^3\,\textrm{s}^{-1}$ ($R_\infty \sim 11 \cdot 10^3\,\textrm{s}^{-1}$).
}
\end{center}\end{figure}

In the case of a qubit encoded into two hyperfine ground states, as for \mg, the finite detuning of the detection laser pulse with respect to the \upstate\ state results in off-resonant excitation of the ion with a rate of approximately
\begin{eqnarray}
\label{eq:depumping_rate}
\frac{\Gamma^3 s_0}{8\Delta_\mindex{HF}^2} \sim 2\pi \times 2.7\,\textrm{kHz} \quad (s_0 \sim 1)\quad,
\end{eqnarray}
where $s_0 = I/I_s$ is the saturation parameter, $I$ the laser intensity and $I_s$ the saturation intensity of the Doppler cooling transition in \mg. Consequently, for long detection times, the ion is off-resonantly pumped into the \downstate\ state, where it enters the cycling transition. Once there, a significantly larger amount of photons is scattered, which leads to a wrong assignment of the ion to the bright state, although it was initially dark. This results in a bias error of the state assignment \cite{Steane:08, Lucas:09, Wineland:07b}, thus imposing a limit on the detection time. The effect of depumping on the number of scattered photons is depicted in \rfig{fig:depumping}\,(a). The average number of detected photons for the bright state is
\begin{eqnarray}
\label{eq:bright_fluorescence}
\bar{\xi}_\ind{B}(\tau) &=& R_\ind{B} \cdot \tau\quad,
\end{eqnarray}
where $R_\ind{B}$ is the scattering rate of the bright state. For the dark state, the depumping effect is taken into account in the following way: If the ion remains in the dark state during the detection interval, only background counts $\bar{\xi}_\infty = R_\infty \cdot \tau$ are measured. Here, $R_\infty$ corresponds to the background scattering rate and $\tau$ to the detection time. In case the ion is depumped at a time $t$ during the detection interval, the number of photons depends on the time ($\bar{\xi}_t = R_\infty t + R_\ind{B}(\tau - t)$ see \cite{Keselman:10} and \rsec{sec:threshold_detection}). Both processes are weighted exponentially by the function $w(t) = \exp(-t/T)/T$, where $T$ denotes the $1/e$-decay time. The average number of detected photons for the dark state then reads
\begin{eqnarray}
\label{eq:dark_fluorescence}
\hspace{-1cm}
\bar{\xi}_\ind{D}(\tau) &=& \int_{0}^{\tau}\!dt w(t) \bar{\xi}_t  + \int_{\tau}^{\infty}\!dt w(t) \bar{\xi}_\infty 
= R_\ind{B} \cdot \tau - \left(R_\ind{B} - R_\infty\right) T \left(1 - e^{-\tau/T}\right).
\end{eqnarray}
The fit of the measured data yields a decay time of $T \approx 61 \pm 4\,\mu$s (see \rfig{fig:depumping}(a)), which is consistent with the expected depumping rate of \req{eq:depumping_rate}.

In principle, the bright state can also be depumped to the dark state if the polarisation of the detection light is not pure. However, this effect can be easily eliminated and will not be considered here. By comparing the initial bright state preparation in the optical pumping cycle with and without the $\sigma$-repumper laser, we typically observe an estimated infidelity of 1-2\,\% in the polarization of the detection light.

One possibility to circumvent the depumping process is to map the hyperfine qubit in a state-selective way onto an optical qubit encoded in a ground and a meta-stable excited optical state. Since the magnitude of the detuning of the detection laser is in the optical regime, off-resonant depumping can be neglected. However, the finite lifetime of the excited meta-stable state also imposes a limitation on the detection time since the ion eventually decays back to the ground state by spontaneously emitting a photon. In addition to that, the finite fidelity of the mapping protocol has to be taken into account \cite{Steane:08}.

\subsubsection{Overlapping distributions.}

In \rfig{fig:depumping}\,(b), the measured probability distributions of the collected fluorescence photons for both bright and dark states of the \mg\ ion in 10\,$\mu$s are shown. As a consequence of the limited detection time and the depumping events, the distributions exhibit a finite overlap. If the number of detected photons lies in the overlapping region, it cannot be unambiguously decided to which state the number of collected photons should be assigned, leading to false state assignments with a bias that depends on the fraction of the photon histograms below (above) the threshold for the bright (dark) state.

\subsubsection{Fluctuations of experimental parameters.} 

Temporal fluctuations in the atomic scattering rate induced by intensity or frequency fluctuations of the detection laser or a change in the bias magnetic field alter the photon distributions of both dark and bright states from shot-to-shot. If the state detection method depends on either form or average value of the photon distribution, the result of the state assignment will be biased by a change in the scattering rate.
 
\subsection{Threshold Detection Technique}
\label{sec:threshold_detection}

In this method, state discrimination is achieved by imposing a threshold $\sigma$ and assigning all detection events which yield a photon number higher than the chosen threshold to the bright state and all events with equal or lower photon numbers to the dark state. The advantage of this method lies in its simplicity and the fact that a single detection cycle suffices to determine the state the ion was projected onto.

\begin{figure}[!h]\begin{center}
	\includegraphics[width=1.0\linewidth]{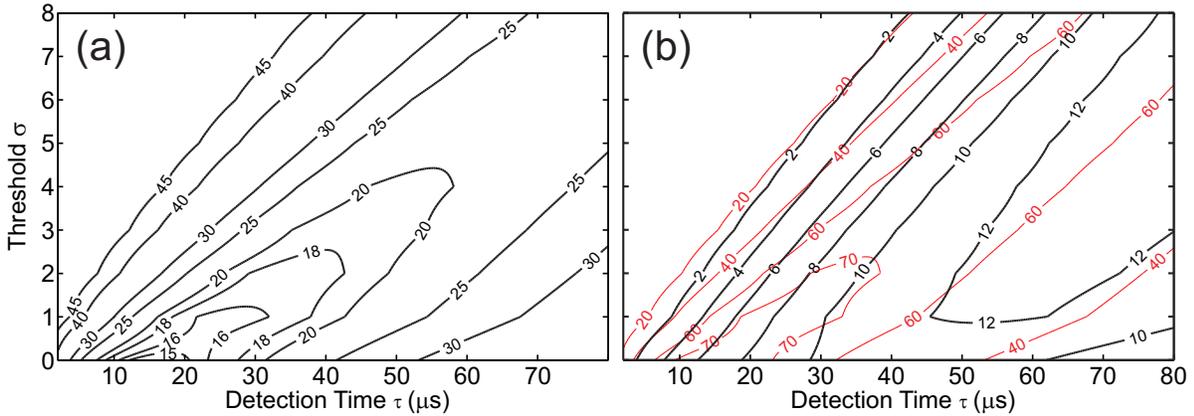}
	\caption{
\label{fig:threshold_detection_error}
Theoretical detection errors in percentage (\%). The scattering rates for both states are $R_\ind{B} \sim 146.3 \cdot 10^3\,\textrm{s}^{-1}$ ($R_\infty \sim 2.9 \cdot 10^3\,\textrm{s}^{-1}$) in this calculation. Part (a) depicts the expected error (\req{eq:threshold_technique}) for the threshold technique, whereas part (b) shows the expected error for the $\pi$-detection ({\it black contour lines}) as a function of the detection time and the threshold (see \rsec{sec:pi-detection} for details). Additionally shown in part (b) is the approximate amount of the remaining statistics (\%) according to \req{eq:pi_det_stat} ({\it red contour lines}). This value reflects the fraction of measurements that are not discarded by the post-selective filtering.
}
\end{center}\end{figure}

The state detection error for the bright state is given by
\begin{eqnarray}
p_\ind{err}^\ind{B} = p( \xi_\ind{B} \le \sigma ) &=& \sum_{k=0}^\sigma \frac{\bar{\xi}_\ind{B}^k e^{-\bar{\xi}_\ind{B}}}{k!}
=
\frac{\Gamma(\sigma + 1, \bar{\xi}_\ind{B})}{\sigma!}\quad;
\end{eqnarray}
For the dark state, it reads in the absence of depumping
\begin{eqnarray}
p_\ind{err}^\ind{D} = p( \xi_\ind{D} > \sigma ) &=& \sum_{k=\sigma+1}^\infty \frac{\bar{\xi}_\infty^k e^{-\bar{\xi}_\infty}}{k!}
= \frac{\gamma(\sigma + 1, \bar{\xi}_\infty)}{\sigma!}\quad,
\end{eqnarray}
where $\Gamma(\sigma, \bar{\xi})$ and $\gamma(\sigma, \bar{\xi})$ are the incomplete upper and lower gamma functions
\begin{eqnarray}
\Gamma(\sigma, \bar{\xi}) = \int_{\bar{\xi}}^\infty dt \, e^{-t}\, t^{\sigma-1}
\quad;\quad
\gamma(\sigma, \bar{\xi}) = \int_0^{\bar{\xi}} dt \, e^{-t}\, t^{\sigma-1}\quad.
\end{eqnarray}
Since depumping is neglected the photon counts of the dark states correspond to the background, i.e.~$\xi_D = \xi_\infty$. The depumping changes only the dark state and is again included by weighting the probability distributions with an exponential decay $w(t)$ again
\begin{eqnarray}
p_\ind{err}^\ind{D} = p( \xi_\ind{D} > \sigma )
&=&
\sum_{k=\sigma+1}^{\infty}
\left[
\int_0^{\tau}\!dt\;w(t) \;
p(k | \bar{\xi}_t)
+
\int_{\tau}^\infty\!dt\;w(t) \;
p(k | \bar{\xi}_\infty)
\right]
.
\end{eqnarray}
The first term describes the case in which a depumping event took place during the detection time $\tau$. This effectively changes the mean value of the Poissonian photon distribution to $\bar{\xi}_t = R_\infty t + R_\ind{B}(\tau - t)$ \cite{Keselman:10}. The second term accounts for the case where only background counts are measured, i.e.~the depumping took place after the detection. The expression can be further simplified to
\begin{eqnarray}
\nonumber
\hspace{-1.5cm}p_\ind{err}^\ind{D} =
\left(
\frac{1}{\beta}
\cdot
e^{-\bar{\xi}_\ind{B}/\beta}
\right)
\cdot
\sum_{k=\sigma+1}^{\infty}
\left[
\frac{\alpha^{-(k+1)}}{k!}
\cdot
\left[
\gamma(k+1, \bar{\xi}_\ind{B}\cdot\alpha)
-
\gamma(k+1, \bar{\xi}_\infty\cdot\alpha)
\right]
\right] +\\
+\quad e^{-\tau/T}
\cdot
\frac{\gamma(\sigma+1, \bar{\xi}_\infty)}{\sigma!}
\quad,
\end{eqnarray}
with the definitions $\beta = (\bar{\xi}_\ind{B} - \bar{\xi}_\infty)\cdot T/\tau$ and $\alpha = 1 - \frac{1}{\beta}$. In \rfig{fig:threshold_detection_error}\,(a), the average detection error
\begin{eqnarray}
\label{eq:threshold_technique}
\epsilon_\ind{th} = \frac{p_\ind{err}^\ind{D} + p_\ind{err}^\ind{B}}{2}
\end{eqnarray}
for the threshold technique is shown as a function of the detection time $\tau$ and the threshold $\sigma$. An increase in the detection time leads to smaller errors for the bright state, but the probability of a depumping event increases, resulting in larger errors for the dark state detection. On the other hand, a higher threshold decreases the dark state error, but larger parts of the bright state photon distribution are wrongly assigned to the dark state. For this reason, there is an optimum detection time for each chosen threshold which minimizes the threshold detection error.

Among the main error sources of the threshold technique is the fact that the bright and dark state photon distributions overlap. We find that replacing the chosen threshold by an interval $(\sigma - \delta, \sigma + \delta)$ which excludes all detection events in the overlapping region does not solve this issue. Instead, in this scenario, the distributions are truncated asymmetrically, especially with an average low number of detected photons, leading to a bias in the assigned state of the ion (see also \rsec{sec:pi-detection}).

\subsection{Distribution-Fit Detection}

\begin{figure}[!h]\begin{center}
	\includegraphics[width=0.6\linewidth]{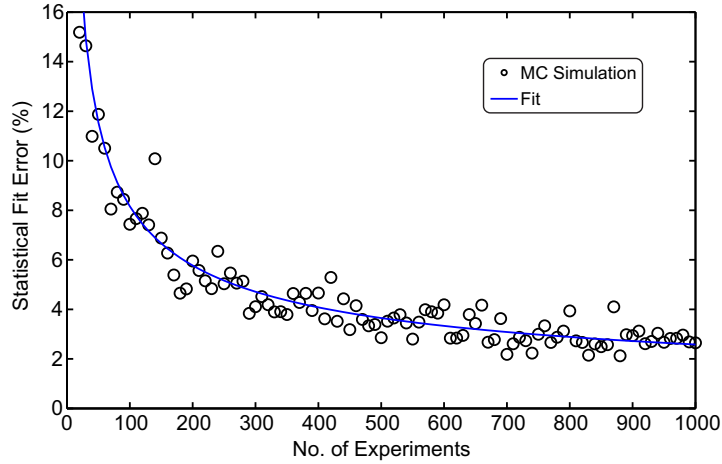}
	\caption{
\label{fig:error_distribution_detection}
Simulated detection error of the distribution-fit technique. The amount of scattered photons is determined in a Monte-Carlo type simulation assuming the ion is in an equal superposition of bright and dark states. The statistical fitting error of the least square fit of the photon histogram to \req{eq:dist_fit} is plotted as a function of the number of experiments. The scattering rates are the same as in \rfig{fig:threshold_detection_error}.
}
\end{center}\end{figure}

In this detection method, the full distribution of the photon histograms of both bright and dark states are taken into account for state discrimination. Prior to the experiments, two calibration photon histograms are taken: one where the ion is prepared in the \downstate\ state by means of optical pumping ($p^\ind{c}(k|\bar{\xi}_\ind{B})$) and a second one where the resonant Doppler-cooling laser is either switched off or the ion is released from the trap and only background counts are observed  ($p^\ind{c}(k|\bar{\xi}_\infty)$) \cite{Hemmerling:11}. Here, $p^\ind{c}(k|\bar{\xi})$ corresponds to the calibrated probability distribution of measuring $k$ photons given an average number of photons of $\bar{\xi}$. The bright state amplitude $a$ is then determined by a least-square fit to a superposition of both distributions
\begin{eqnarray}
\label{eq:dist_fit}
p(k | a) = a \cdot p^\ind{c}(k|\bar{\xi}_\ind{B}) + (1-a) \cdot p^\ind{c}(k|\bar{\xi}_\infty)\quad.
\end{eqnarray}
The expected detection error of this {\it distribution-fit detection} can be determined by Monte-Carlo simulations and is shown in \rfig{fig:error_distribution_detection} as a function of the number of total detection events. As postulated by the law of large numbers, the error decreases with the square-root of the number of experiments. For our experimental parameters of typical 300-1000 experiments, a value of 2-4\% is expected, which is consistent with the standard deviation of the measurement result using the distribution-fit technique.

While the predicted error under the assumption of our typical experimental parameters is smaller than the error of the threshold detection, a sufficient number of measurements is required to fit the photon distributions. This detection method does not provide single-shot results.

\subsection{$\pi$-Detection}
\label{sec:pi-detection}

\begin{figure}[!h]\begin{center}
	\includegraphics[width=0.50\linewidth]{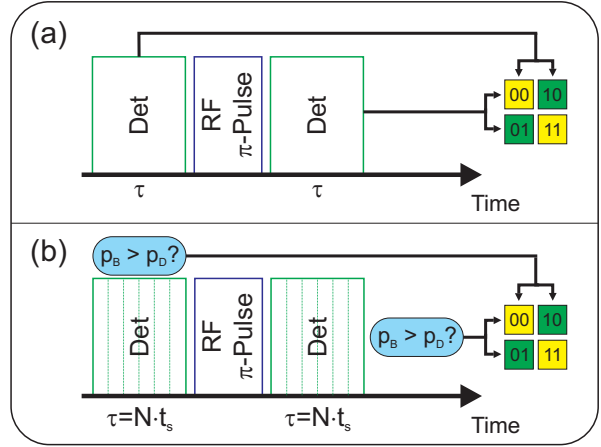}
	\caption{
\label{fig:pi_detection_scheme}
(a) $\pi$-detection scheme. Two identical threshold detection events are combined with an interleaved radio-frequency induced spin-flip. The detection result is only accepted if both detection events yield opposite results ({\it green boxes}). In the case of identical results, the event is discarded ({\it yellow boxes}). (b) $\pi$-Bayesian detection. Similar to the $\pi$-detection, two Bayesian detection events are connected with an intermediate spin-flip. Here, a single Bayesian detection event consists of a $N$ consecutive sub-bins of length $t_s$. The detection result of each pulse is determined by the maximum likelihood formula (see e.g.~\req{eq:maximum_likelihood}).
}
\end{center}\end{figure}

In order to overcome the described limitations of the previous methods, two detection events which use the threshold technique are combined with an intermediate spin-flip and anti-correlations of both detection results are analyzed. The detection pulse sequence is shown in \rfig{fig:pi_detection_scheme}\,(a). If both events yield the same state result, the detection is identified as false and discarded. For instance, assume the ion is in the bright state before detection, but is assigned to the dark state since the number of detected photons is below the threshold. The spin-flip brings the ion into the dark state where it is detected for a second time. If this measurement (correctly) yields a dark state again, the whole detection process is discarded, since a false detection was recognized. Thus, the {\it $\pi$-detection method} effectively acts as a statistical filter for the measured data.

\begin{figure}[!h]\begin{center}
	\includegraphics[width=1.0\linewidth]{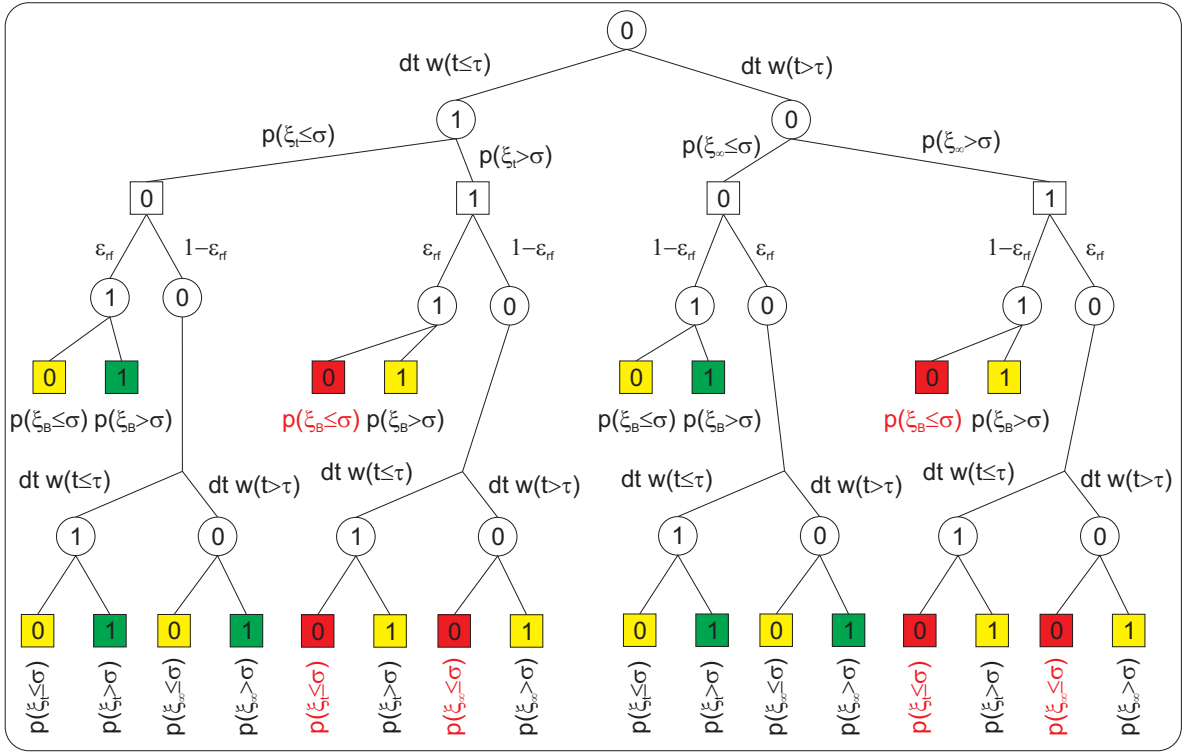}
	\caption{
\label{fig:pi_detection_tree_zero.eps}
Decision tree for $\pi$-detection. Assuming the ion to be initially in the dark state, all possible paths of both detection events with their corresponding probabilities are shown. Also included is a possible error in the spin-flip by assuming a finite fidelity of $1-\epsilon_\ind{rf}$ for a successful inversion. The first branches are marked for clarity: I) Off-resonant excitation II) Threshold detection III) Spin-flip. Circles correspond to the actual state of the ion, whereas rectangular boxes denote the detection result (1 for the bright and 0 for the dark state). The color coding distinguishes between correct detection events ({\it green}), discarded events ({\it yellow}) and detection errors ({\it red}).
}
\end{center}\end{figure}

The detection error for this scheme is determined by combining the errors of two threshold detection events and a finite fidelity of the intermediate spin-flip of $1-\epsilon_\ind{rf}$, while neglecting discarded events. All possible paths of the whole detection process are shown in \rfig{fig:pi_detection_tree_zero.eps}, assuming that the ion is initially in the dark state. The decision tree for the bright state is similar. Summing over all possible events yields for the dark state error including depumping effects
\begin{eqnarray}
&&p_\ind{$\pi$, err}^\ind{D} =\\
\nonumber
&&
\left(\int_0^{\tau}\!dt\;w(t) \;
p(k>\sigma | \bar{\xi}_t)\right)
\left\{
p(\xi_\ind{D} \le \sigma)
+
\epsilon_\ind{rf}
\left(
p(\xi_\ind{B} \le \sigma) - p(\xi_\ind{D} \le \sigma)
\right)
\right\}\\
\nonumber
&&+
\left(\int_{\tau}^{\infty}\!dt\;w(t) \;
p(k>\sigma | \bar{\xi}_\infty)\right)
\left\{
p(\xi_\ind{B} \le \sigma)
-
\epsilon_\ind{rf}
\left(
p(\xi_\ind{B} \le \sigma) - p(\xi_\ind{D} \le \sigma)
\right)
\right\}\\
\nonumber
&&
=
\frac{1}{\beta} e^{-\bar{\xi}_\ind{B}/\beta}
\sum_{k=\sigma+1}^{\infty}
\left[
\frac{\alpha^{-(k+1)}}{k!}
\cdot
\left[
\gamma(k+1, \bar{\xi}_\ind{B} \cdot \alpha)
-
\gamma(k+1, \bar{\xi}_\infty \cdot \alpha)
\right]
\right]\\
\nonumber
&&
\hspace{1cm}\cdot
\left\{
p(\xi_\ind{D} \le \sigma)
+
\epsilon_\ind{rf}
\left(
p(\xi_\ind{B} \le \sigma) - p(\xi_\ind{D} \le \sigma)
\right)
\right\}\\
\nonumber
&&
+
e^{-\tau/T}
\cdot
\frac{\gamma(\sigma+1, \bar{\xi}_\infty)}{\sigma!}
\left\{
p(\xi_\ind{B} \le \sigma)
-
\epsilon_\ind{rf}
\left(
p(\xi_\ind{B} \le \sigma) - p(\xi_\ind{D} \le \sigma)
\right)
\right\}\quad.
\end{eqnarray}
On the other hand, if the ion is initially in the bright state, the error is given by
\begin{eqnarray}
p_\ind{$\pi$, err}^\ind{B}
&=&
p(\xi_\ind{B} \le \sigma) \cdot
\left\{
p(\xi_\ind{D} > \sigma)
+
\epsilon_\ind{rf} \cdot
\left(
p(\xi_\ind{B} > \sigma) - p(\xi_\ind{D} > \sigma)
\right)
\right\}\quad.
\end{eqnarray}
The advantage of this method is that two small probabilities are multiplied, yielding an overall smaller error than the pure threshold detection method. In \rfig{fig:threshold_detection_error}\,(b), the average detection error of bright and dark states as a function of the chosen threshold $\sigma$ and the detection time $\tau$ is shown. While the error is almost an order of magnitude smaller than for the threshold technique and tends to zero for very small detection times, it should be noted that this is at the expense of the usable data, also shown in the figure.
For this calculation, the intermediate radio-frequency transition was assumed to have a fidelity of $98\%$, which is a conservative lower bound of the actual fidelity \cite{Wineland:11}. If necessary, the implementation of different, more robust techniques that are independent of the area of the radio-frequency pulse, such as rapid adiabatic fast passage \cite{Schmidt-Kaler:05} or STIRAP \cite{Shore:98} can improve the fidelity even further.

\begin{figure}[!h]\begin{center}
	\includegraphics[width=0.75\linewidth]{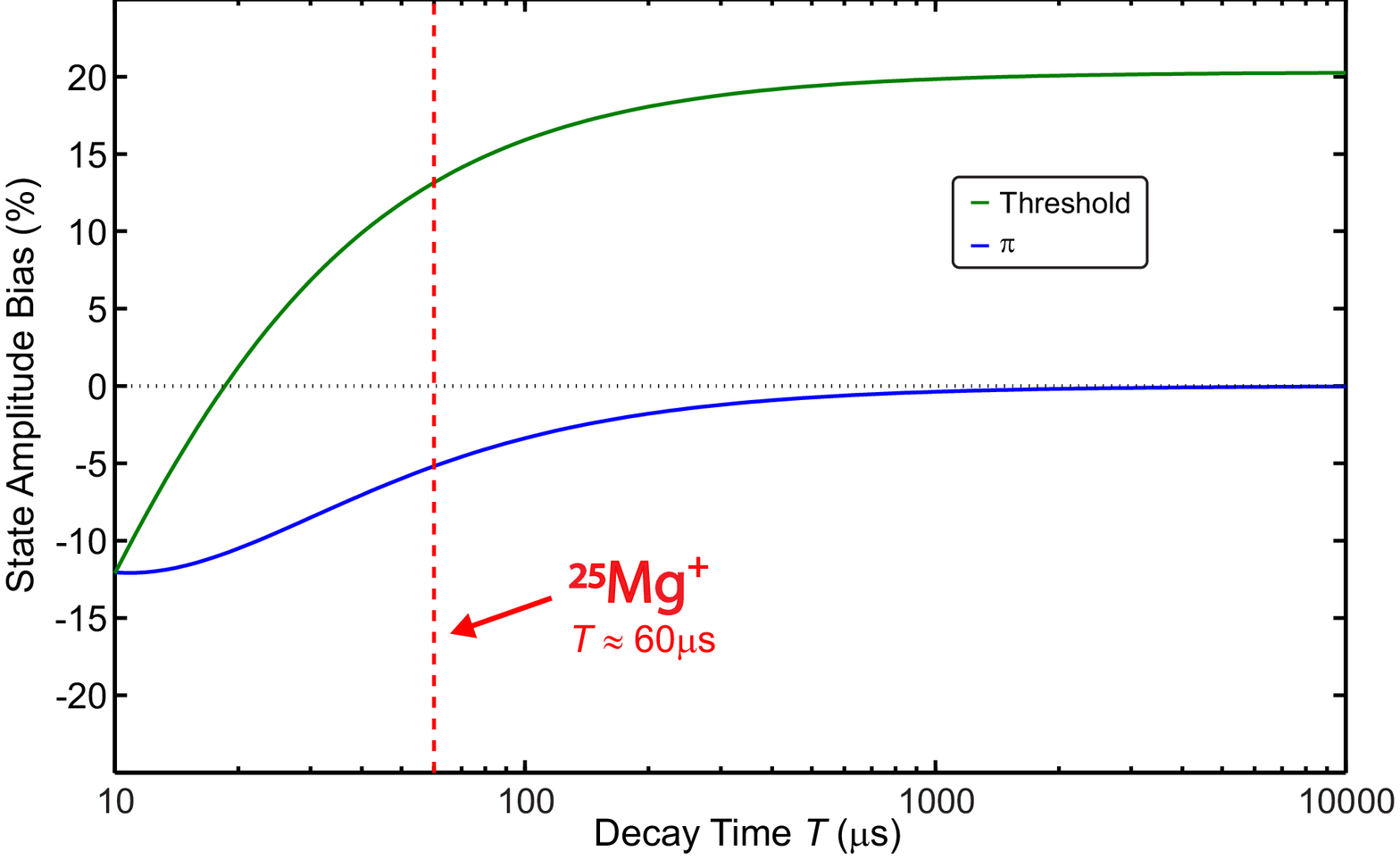}
	\caption{
\label{fig:th_pi_detection_bias}
Calculated bias $b_\ind{th,$\pi$}$ of the detection methods as a function of the decay time $T$. The {\it black dotted} line is a guide to the eye and corresponds to the zero bias line. The {\it red dashed} line reflects the case of the \mg\, system with a decay time of approximately $T \approx 60\,\mu$s. The calculation was done for a detection time of $10\,\mu$s and a threshold of $\sigma = 0$. The scattering rates $R_\ind{B,D}$ are the same as in \rfig{fig:threshold_detection_error}.
}
\end{center}\end{figure}

While the post-selective analysis for the $\pi$-detection method reduces the detection error by discarding all photon counting events in the overlapping region of the histograms, events which undergo depumping during detection introduce a remaining bias in the resulting state amplitude. This type of error is already taken into account in the above calculation, but shall be analysed here in more detail. As described in \rsec{sec:threshold_detection}, an asymmetric discarding of events with respect to both the bright and the dark states will introduce a bias towards the state with less discarded events. This bias is thus readily determined by comparing the difference in the probabilities of correctly detecting either the bright or the dark state. Neglecting the infidelity of the interleaved spin-flip, it reads
\begin{eqnarray}
b_\ind{th} = p(\xi_\ind{D} \le \sigma) - p(\xi_\ind{B} > \sigma)\quad.
\end{eqnarray}
In the case of the $\pi$-detection, a sum over all possible correct detection events yields for the bias
\begin{eqnarray}
\nonumber
b_\ind{$\pi$} &=& 
p(\xi_\ind{B} > \sigma) \cdot \left(\int_{\tau}^{\infty}\!dt\;w(t) \;
p(k \le \sigma | \bar{\xi}_\infty)\right)\\
\nonumber &&+
p(\xi_\ind{D} > \sigma) \cdot \left(\int_0^{\tau}\!dt\;w(t) \;
p(k \le \sigma | \bar{\xi}_t)\right)\\
&&-
p(\xi_\ind{D} \le \sigma) \cdot p(\xi_\ind{B} > \sigma)\quad.
\end{eqnarray}
Both biases $b_\ind{th}$ and $b_\ind{$\pi$}$ are plotted in \rfig{fig:th_pi_detection_bias} as a function of the decay time $T$. Evidently, the threshold method introduces a systematic bias to the measurement result. The bias is shifted towards the dark state with increasing decay time and levels at approximately 20\,\% for $T \rightarrow \infty$, for the parameters described previously. In contrast to that, the $\pi$-detection method operates at a smaller bias and culminates in a bias-free detection for systems with large decay times. This is a result of the symmetric approach of comparing two consecutive and inverted detection events. The decay time of the \mg\ qubit system being approximately $T \approx 60\,\mu$s, the $\pi$-detection method is estimated to have a bias of $-5.2\,\%$, whereas the bias of the conventional threshold method is estimated to be $13.1\,\%$.

\subsection{Bayesian and $\pi$-Bayesian Detections}

In the previously discussed methods, the collected number of photons integrated over the detection time $\tau$ was considered. This neglects all information which could be gained from the arrival times of the photons. This information can be included by dividing the detection interval into $N$ sub-intervals of length $t_s$. The state of the ion is then inferred from the time-resolved series of detected photons $\{\xi_i\}$, where $i = 1, \ldots, N$. The likelihoods $p(\{\xi_i\}|\textrm{B,D})$ of observing a particular series of photons given that the ion is in either bright (B) or dark states (D) are calculated and the ion is assigned to the state with the highest likelihood \cite{Steane:08, Wineland:07b}.

The likelihood of the photon series originating from the bright state is given by
\begin{eqnarray}
\label{eq:maximum_likelihood}
p(\{\xi_i\} | \textrm{B})
=
\prod_{k=0}^{N} p(\xi_i|\bar{\xi}_\ind{B})\quad.
\end{eqnarray}

In case of the dark state, it reads
\begin{eqnarray}
\nonumber
p(\{\xi_i\} | \textrm{D}) =
\left[
\left( 1 - e^{-t_s/T} \right)
\sum_{j=1}^{N}
e^{-(j-1)t_s/T}
\prod_{k=1}^{j-1}
p(\xi_k|\bar{\xi}_\infty)
\prod_{l=j}^{N}
p(\xi_l|\bar{\xi}_\ind{B})
\right] +\\
\hspace{3cm}
\left[
e^{-Nt_s/T}
\prod_{j=1}^N
p(\xi_j|\bar{\xi}_\infty)
\right]
\quad,
\end{eqnarray}
where it was assumed that depumping events only happen at the end of a sub-interval. The detection error is determined by Bayes' theorem, i.e.~the probability that the bright state is wrongly deduced to be dark reads \cite{Steane:08}
\begin{eqnarray}
p_\ind{Bay, err}^\ind{B}
=
1 - p(\textrm{D} | \{\xi_i\})
=
p(\textrm{B} | \{\xi_i\})
=
\frac{ p(\{\xi_i\} | \textrm{B})  }{ p(\{\xi_i\} | \textrm{D}) + p(\{\xi_i\} | \textrm{B}) }\quad,
\end{eqnarray}
and vice versa for the dark state. This can be used to determine the detection error in real time and stop the measurement as soon as the desired error level is reached \cite{Wineland:07b}.

In an identical way as for the threshold detection method, the Bayesian detection can be extended to a $\pi$-Bayesian detection, as shown in \rfig{fig:pi_detection_scheme}\,(b). Here, two detection events which use the maximum likelihood to determine the state of the ion are combined with an intermediate spin-flip. This again serves as a statistical filter and discards correlated events. The detection error is similarly calculated. For instance, while neglecting the imperfect fidelity of the spin-flip, the error of the bright state detection reads
\begin{eqnarray}
p_\ind{$\pi$-Bay, err}^B(\{\xi^{(1)}_i, \xi^{(2)}_j\})
=
p(\textrm{D}|\{\xi^{(1)}_i\})
\cdot
p(\textrm{B}|\{\xi^{(2)}_j\})\quad.
\end{eqnarray}

\begin{figure}[!h]\begin{center}
	\includegraphics[width=1.0\linewidth]{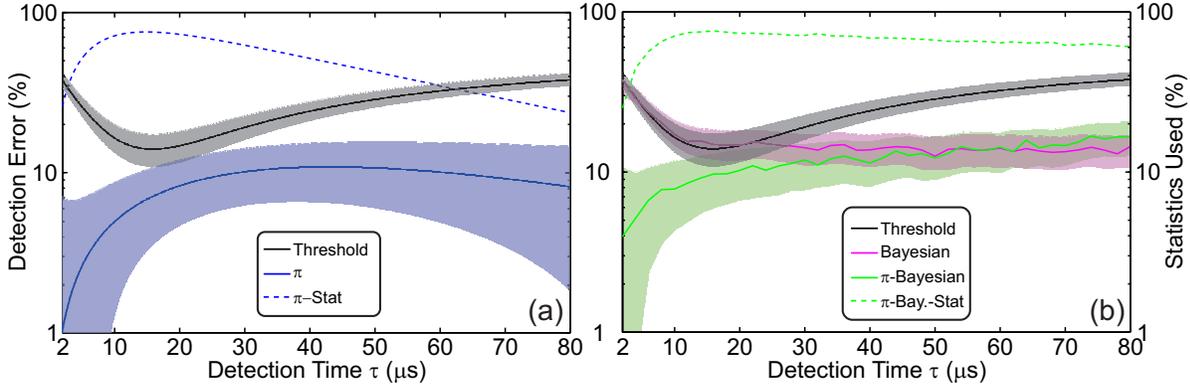}
	\caption{
\label{fig:det_time_err}
Theoretical error for different detection methods as a function of the detection time. The parameters are the same as in \rfig{fig:threshold_detection_error}. For the Bayesian method, a Monte-Carlo simulation with 2000 events at each point was used to determine the error. The dashed lines correspond to the amount of remaining statistics for the $\pi$-type methods. A threshold $\sigma = 0$ is chosen. The shaded areas reflect the expected additional statistical error for 1000 experiments. (a) Threshold and $\pi$-detection methods. (b) Threshold, Bayesian and $\pi$-Bayesian methods. For the simulation of the Bayesian techniques, a sub-bin time of $2\,\mu$s was used.
}
\end{center}\end{figure}

In \rfig{fig:det_time_err}\,(a), calculations of the average detection error $\epsilon_\ind{th, $\pi$}$ of both the threshold and $\pi$-detection are shown in comparison as a function of the detection time, illustrating the effect of depumping on the detection error. While there is an optimal point for the threshold technique due to an increase in error from depumping for longer detection times, the $\pi$-detection yields smaller single-shot errors at the expense of the number of used events, as shown by the dashed blue curve. Consequently, the statistical error which results from the quantum projection noise increases as the number of used events decreases \cite{Wineland:93}. This is shown by the plotting the $95\%$ confidence interval $\epsilon_\ind{th, $\pi$} \pm 1.96 \sqrt{a(1-a)/n}$ (shaded areas in the plots), where $a$ represents the bright state amplitude, similar to \req{eq:dist_fit}. The number of experiments is $n=1000$ for the threshold technique and is scaled by the percentage of used events for the $\pi$-techniques. Furthermore, we use $a=0.5$ as a conservative estimate, owing to the maximal quantum projection noise if the ion is in an equal superposition of bright and dark state.

Although the $\pi$-detection discards many events, the resulting statistical spread still yields a smaller overall detection error than the threshold technique. Nevertheless, it should be noted that applying the $\pi$-detection in a quantum algorithm with several consecutive detection steps, the algorithmic sequence needs to be repeated in case of a single inconclusive result. It is worthwhile mentioning that the non-zero single-shot error of the threshold technique leads to a bias that fundamentally does not average to zero, whereas the statistical uncertainty can be improved by increasing the number of measurements.
 
The same holds for the Bayesian techniques. In \rfig{fig:det_time_err}\,(b), Monte-Carlo simulations of both Bayesian and $\pi$-Bayesian technique are shown in comparison. The threshold technique yields similar results to the Bayesian method for short detection times. The observation of the photon series only improves for detection times longer than 20\,$\mu$s. Applying the $\pi$-method to the Bayesian scheme additionally decreases the detection error for short detection times at the expense of the used events.

Choosing an optimal threshold for each detection time for both threshold and $\pi$-detection decreases the error for both methods further. Although smaller single-shot errors are achieved for longer detection times with an optimal threshold, there is a trade off by discarding more events with the $\pi$-technique.

\section{Robustness and Sensitivity of Detection Schemes}

A major difference among the detection methods is their sensitivity to the chosen threshold and to fluctuations in the atomic scattering rate from technical noise. While the Bayesian and distribution-fit detection methods are inherently threshold free, the threshold and $\pi$-detection schemes may be affected by choosing the wrong threshold. This is discussed in detail in \rapp{app:threshold_sensitivity}.

Fluctuations of experimental parameters, such as the power and frequency of the employed detection laser or the strength of the external magnetic field, are a major experimental challenge since they change the fluorescence rate of the atom and introduce further errors into the measurement process. Thus, it is highly desirable to aim for robust detection methods. Here, sensitivity with respect to power fluctuations was measured for the different detection techniques.

\begin{figure}[!h]\begin{center}
	\includegraphics[width=0.75\linewidth]{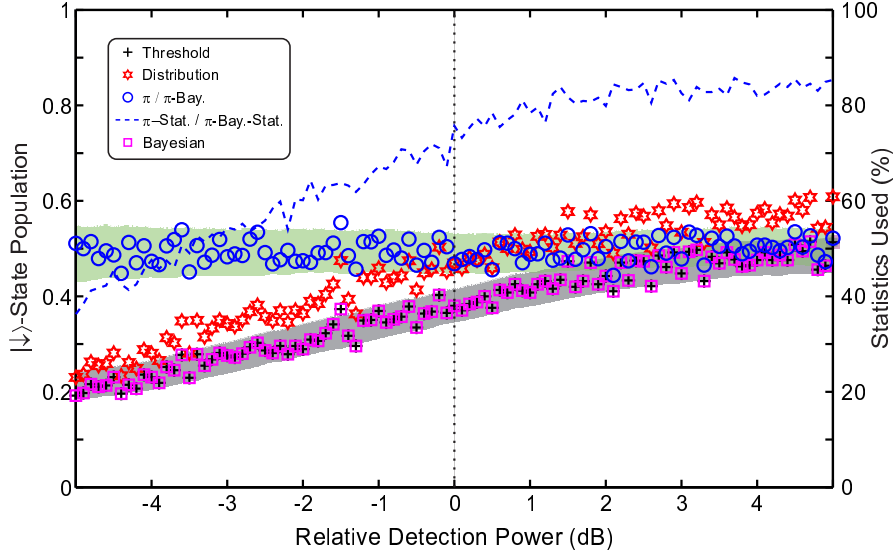}
	\caption{
\label{fig:power_fluctuations}
Robustness with respect to power fluctuations for all detection methods. The ion is initialized in an equal superposition of bright and dark states and the nominal detection power is 0\,dB, corresponding to the saturation intensity. Detection methods that depend on the threshold experience large systematic deviations from the initial state result. Both $\pi$-detection techniques are almost independent of the laser power at the expense of usable events. Shaded areas depict the statistical error as in \rfig{fig:det_time_err}. The detection time for the threshold detection and both threshold events of the $\pi$-detection was $\tau = 10\,\mu$s. In the case of the Bayesian detections five sub-bins with a length $t_s = 2\,\mu$s were used.}
\end{center}\end{figure}

In order to simulate power fluctuations, the ion was initialized in an equal superposition of the \downstate\ and \upstate\ states by means of a $\pi/2$ radio-frequency pulse. The power of the detection laser was scanned over $\pm 5$\,dB around its optimal value of 0\,dB (corresponding to saturation intensity), as shown in \rfig{fig:power_fluctuations}. At each point, $3 \times 250$ measurements were taken. Since a change in laser power alters the photon distributions, one expects the detection methods that depend on the threshold to be especially sensitive towards power changes. This is confirmed in the experiment. While the threshold, distribution-fit and Bayesian techniques tend to over- and underestimate the actual \downstate\ population of the ion by 30\%, both $\pi$-detection techniques show no significant dependency on power fluctuations, since wrongly assigned events are systematically discarded. Given a typical detection time of $10\,\mu$s, the Bayesian and the threshold techniques yield comparable results, which is consistent with the fact that their individually expected error is similar, as shown in \rfig{fig:det_time_err}\,(b). Additionally, the expected statistical error due to the finite number of measurements, similarly to \rfig{fig:det_time_err}, is indicated by the shaded areas. A linear fit to the interval between -1\,dB and +1\,dB yields the following changes of the amplitude for the different techniques: (a) threshold technique: 0.$032 \pm 0.011$/dB, (b) $\pi$-detection: $-0.009 \pm 0.016$/dB and (c) distribution-fit technique: $0.037 \pm 0.015$/dB.

It is important to stress that the systematic bias found in the conventional detection methods dominates the overall error in comparison to the statistical error. Although the statistical error can be decreased by obtaining more measurements, the systematic bias is only corrected in the $\pi$-detection methods.

\section{Conclusion}

We demonstrated a novel qubit detection technique which combines two detection events with an intermediate state inversion. The expected anti-correlation of both detection events is used in a post-selective statistical filtering. As a consequence, a higher detection fidelity is achieved at the expense of the number of usable events. Although this results in an increase of the statistical error due to quantum projection noise, the systematic bias error of the conventional threshold technique is in comparison significantly larger and does not average to zero. We compared the $\pi$-detection to other methods in terms of their sensitivity to fluctuations of experimental parameters. In particular, the $\pi$-detection method shows no significant dependence on the detection time, the chosen threshold or power fluctuations of the detection laser, which makes it a rather robust experimental tool.

Furthermore, especially in a regime of low photon count rate, the overlap of the photon histograms for bright and dark state imposes a limitation on the achievable contrast of Rabi oscillations to the conventional threshold technique, whereas almost full contrast is observed using the $\pi$-detection (see \rapp{app:threshold_sensitivity}).

Another advantage of this method is its simplicity, requiring only an rf-induced state inversion for qubits encoded in hyperfine ground states. This technique applied to optically encoded qubits would require a narrow-bandwidth laser for coherent manipulation, which is often available in ion trap quantum computing and precision spectroscopy experiments.

The $\pi$-detection method could also be applied to multiple qubit systems where the qubits are individually accessible for detection, as e.g.~in ion traps where multiple ions are spatially resolved \cite{Wineland:08}. Furthermore, the concept of this detection scheme is very general and can be applied to a multitude of systems for which non-destructive detection methods can be implemented, such as neutral atoms in deep optical lattices \cite{Greiner:09, Bloch:10}, measurements of the state-dependent reflection and transmission signals of molecules or atoms in cavities \cite{Zoller:06, Reichel:10}, and nitrogen vacancy centers \cite{Lukin:07, Lukin:08}.

\ack

We would like to thank Rainer Blatt for generous loan of equipment and support. We further acknowledge the technical support of P.-C. Carstens and S. Klitzing. We gratefully acknowledge financial support by the Austrian START program of the Austrian Ministry of Education and Science, the Cluster of Excellence QUEST, Hannover, and the Physikalisch-Technische Bundesanstalt, Braunschweig. We would like to thank D Leibfried for helpful comments on the manuscript.

\begin{appendix}

\section{Threshold Sensitivity}
\label{app:threshold_sensitivity}

\begin{figure}[!h]\begin{center}
	\includegraphics[width=0.9\linewidth]{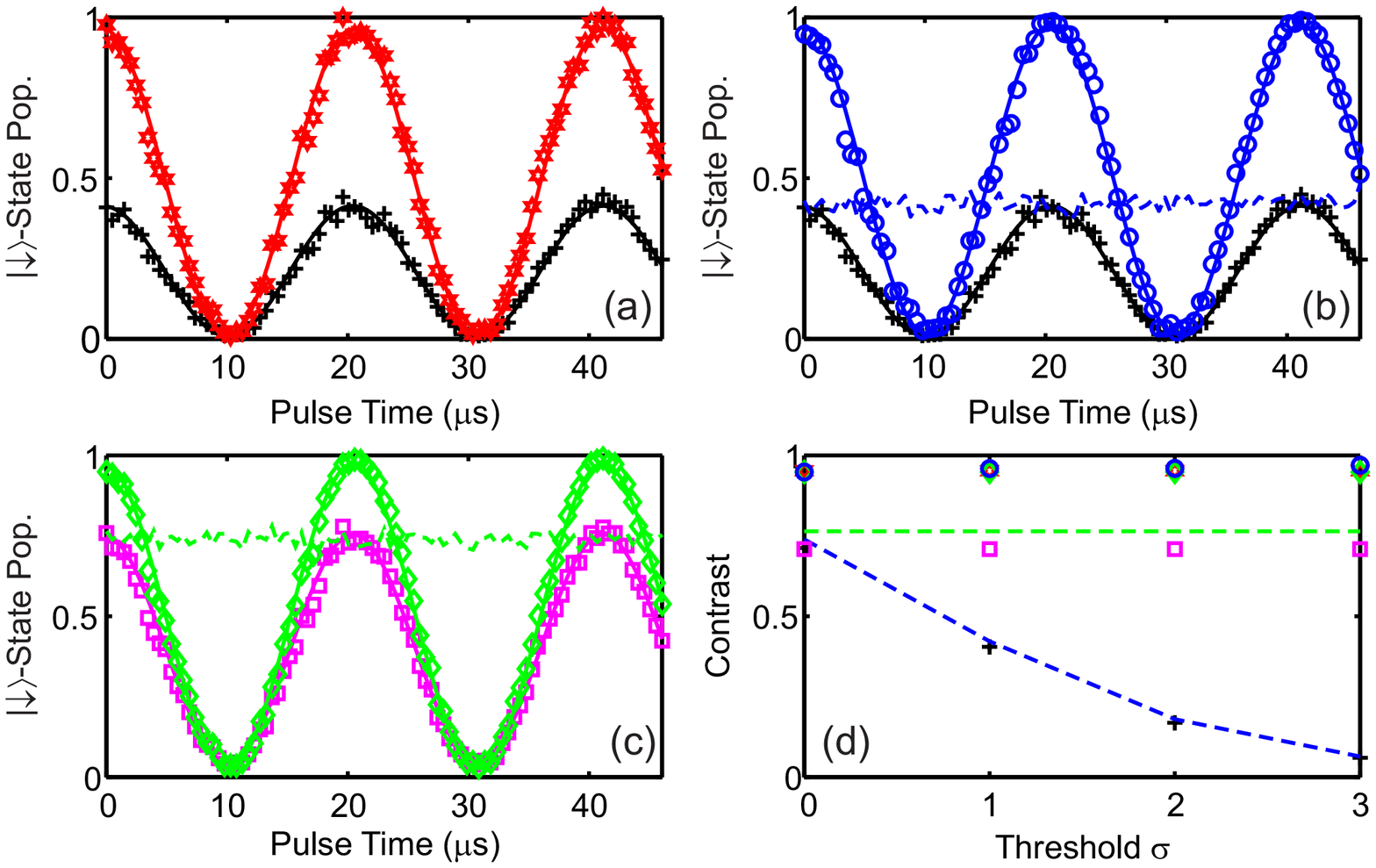}
	\caption{
\label{fig:rabi_flops}
Experimental rabi oscillations analyzed with all detection schemes. Part (a) shows the result of the threshold detection for a threshold of 1 photon ({\it black crosses}) (fitted contrast: $41\%$) and the distribution-fit technique ({\it red stars}) ($95\%$). Part (b) depicts the result of the threshold and the $\pi$-detection technique for a threshold of 1 photon ({\it blue circles}) ($96\%$). In addition to that, the percentage of remaining statistics is plotted ({\it blue dashed line}). Part (c) depicts the result of the Bayesian ({\it magenta boxes}) ($71\%$) and the $\pi$-Bayesian detection ({\it green diamonds}) ($95\%$), along with the amount of statistics used ({\it green dashed line}). Each point is the result of averaging over $3 \times 250$ experiments. The estimated statistical errors are on the order of 3-4\% and are omitted for clarity. Here, the detection time for the threshold detection and both threshold events of the $\pi$-detection was $\tau = 10\,\mu$s. In the case of the Bayesian detections five sub-bins with a length of $t_s = 2\,\mu$s were used. (d) shows the fitted contrast of the Rabi oscillation as a function of the chosen threshold with the same color coding as in parts (a) - (c).
}
\end{center}\end{figure}

We investigated the sensitivity of the detection methods to the threshold by observing the contrast of radio-frequency induced Rabi oscillations. \rfig{fig:rabi_flops} shows a single experimental data set analyzed with different detection techniques. In subplot (a), Rabi oscillations are shown for the threshold and the distribution-fit method for a threshold of 1 photon. Clearly, the contrast resulting from the threshold technique is degraded, owing to the non-optimal threshold; subplot (b) depicts the analysis of the data with the $\pi$-detection compared to the threshold method for the same threshold of 1 photon. The $\pi$-detection yields almost unity contrast by discarding over $\!50\%$ of the events. These correspond to the events that were wrongly assigned to the dark state by the threshold method. This effect can be understood by considering the amount of statistics used by the $\pi$-detection. Neglecting the fidelity of the radio-frequency and the depumping effect for simplicity, the percentage of used statistics is given by
\begin{eqnarray}
\label{eq:pi_det_stat}
p_\ind{$\pi$, stat} = 
1
-
p(\xi_\ind{B} \le \sigma)
-
p(\xi_\infty > \sigma)
+
2
\cdot
p(\xi_\ind{B} \le \sigma)
\cdot
p(\xi_\infty > \sigma)\quad.
\end{eqnarray}
In a regime where the threshold is so high that the error of the dark state detection is negligible, the remaining statistics correspond to $p_\ind{$\pi$, stat} \approx 1 - p(\xi_\ind{B} \le \sigma)$. On the other hand, applying the threshold technique in this regime, the amplitude of the Rabi oscillations is determined by the error of the bright detection, i.e.~$A_\ind{th} = p_\ind{$\pi$, stat}$ which is confirmed by the measurements in \rfig{fig:rabi_flops}\,(b).

Similarly, the $\pi$-Bayesian detection filters correlated events and yields an almost unity contrast as opposed to the pure Bayesian detection. Both are depicted in \rfig{fig:rabi_flops}\,(c). In subplot (d), the dependence of all methods is summarized by plotting the fitted contrast of the Rabi oscillations as a function of the threshold. While the distribution and both Bayesian techniques are fundamentally threshold free, the threshold method experiences a strong decrease of the contrast with increasing threshold. On the other hand, the $\pi$-detection discards many events and the resulting contrast is almost independent of the threshold.

\end{appendix}

\section*{References}
\bibliographystyle{h-physrev3}
\bibliography{Boerge_Hemmerling_A_Novel_Robust_Quantum_Detection_Scheme_Literatur}

\end{document}